\documentclass[a4paper,11pt]{article}
\pdfoutput=1 

\usepackage{jheppub} 
\usepackage{slashed}

\usepackage[T1]{fontenc} 

\title{\boldmath Spin $ 2 $ spectrum for marginal deformations of 4d $ \mathcal{N}=2 $ SCFTs}

\author[a]{Sourav Roychowdhury}
\author[b]{and Dibakar Roychowdhury}

\affiliation[a]{School of Physical Sciences, Indian Association for Cultivation of Science, \\Kolkata 700032, West Bengal, India}
\affiliation[b]{Department of Physics, Indian Institute of Technology Roorkee, \\Roorkee 247667, Uttarakhand, India }

\emailAdd{spssrc2727@iacs.res.in}
\emailAdd{dibakar.roychowdhury@ph.iitr.ac.in}
\abstract{We compute spin $ 2 $ spectrum associated with massive graviton fluctuations in $\gamma$-deformed Gaiotto-Maldacena background those are holographically dual to marginal deformations of $\mathcal{N}=2$ SCFTs in four dimensions. Under the special circumstances, we analytically estimate the spectra both for the $ \gamma $- deformed Abelian T dual (ATD) as well as the non-Abelian T dual (NATD) cases where we retain ourselves upto leading order in the deformation parameter. Our analysis reveals a continuous spectra which is associated with the breaking of the $ U(1) $ isometry (along the directions of the internal manifold) in the presence of the $ \gamma $- deformation. We also comment on the effects of adding flavour branes into the picture and the nature of the associated spin $ 2 $ operators in the dual $ \mathcal{N}=1 $ SCFTs.}
\begin{document} 
\maketitle
\flushbottom
\section{Introduction and summary}
$ \mathcal{N}=2 $ SCFTs \cite{Gaiotto:2009we} those live in four dimensions and their realization as dual supergravity solutions \cite{Gaiotto:2009gz}-\cite{Petropoulos:2014rva} has witnessed as steady progress over the past one and half decade. In particular, the realization of some of these Type IIA geometries \cite{Reid-Edwards:2010vpm}-\cite{Aharony:2012tz} (also known as the Gaiotto-Maldacena background) as Abelian T dual (ATD) as well as non-Abelian T dual (NATD) solutions of $ AdS_5 \times S^5 $ has generated renewed interest in the recent years \cite{Sfetsos:2010uq}-\cite{Roychowdhury:2021eas}.

In a typical Hanany-Witten set up \cite{Hanany:1996ie}, the $ \mathcal{N}=2 $ dualities can be realised as an intersection of NS5-D4-D6 branes whose low energy (IR) description is characterised by a quiver gauge theory containing both color ($ N_c $) as well as flavour ($ N_f $) nodes. The color degrees of freedom are sourced due to D4 branes while the flavour modes come through D6s.

In the IR, the quiver is represented by a (super)conformal fixed point which preserves a global $ SO(2,4)\times SU(2)_{\chi, \xi} \times U(1)_{\beta} $ symmetry. The holographic limit is that for which all gauge nodes become large, $ N_c \rightarrow \infty $. This is the limit in which the dual supergravity description makes sense.

In the present paper, however, we are interested in certain holographic aspects associated to \emph{marginal} deformations of $ \mathcal{N}=2 $ SCFTs those were introduced recently by authors in \cite{Nunez:2019gbg}. The corresponding quiver is conjectured to be describing a superconformal fixed point with $ \mathcal{N} =1$ supersymmetries. The dual supergravity ($ \gamma $- deformed) background can be obtained by applying an $ SL(3,R) $ transformation \cite{Lunin:2005jy}-\cite{Gauntlett:2005jb} in the eleven dimensional description and thereby following a Type IIA reduction.

The purpose of the present paper is to explore the properties of spin 2 operators associated to $ \mathcal{N}=1 $ superconformal quivers by probing the structure of linearised gravitational perturbations in the dual Type IIA set up. The spin 2 spectrum is what is identified through the massive graviton equation \cite{Passias:2018swc}-\cite{Rigatos:2022ktp} in the dual supergravity description.

The linearised perturbation equation is in principle difficult to solve analytically, except for certain special choices for the associated quantum numbers. In fact, we are able to solve these equations for a particular choice of the quantum number $m=0$ that is associated with the $U(1)_{\beta} $ isometrty of the parent Type IIA background.
It turns out that the corresponding graviton equation is analytically tractable which yields a regular solution. On the other hand, for $m\neq0$, the corresponding graviton equation results into a divergent solution which is therefore discarded.

Upon fixing the quantum number $m=0$, we solve the massive graviton equation considering the effects of the $ \gamma $- deformation upto leading order in the perturbative expansion. Contrary to the undeformed case \cite{Itsios:2019yzp}, we notice that the effect of the $ \gamma $- deformation is to change the spectrum from discrete to continuous. We identify this as an artefact of the breaking of the spherical symmetry $ S^2(\chi ,\xi) $ in the presence of the $ \gamma $- deformation.

 These modes are associated to the $ \psi $- direction of the internal manifold which could be recast as an isometric direction for the $ \gamma $-deformed ATD \footnote{One way to realise this isometry for the ATD in Type IIA is through a suitable choice of gauge. We elaborate more on this as we progress in section 4 (see Footnote 3).}. However, one looses this isometry for the $ \gamma $- deformed NATD. One might expect a different scenario once the flavour D6 branes are placed so that the $ \psi $- direction is bounded. However, as we outline in the Appendix \ref{appen c}, the corresponding massive graviton equations are quite nontrivial to deal with and specially in the presence of the $ \gamma $- deformation.

The rest of the paper is organised as follows. We briefly outline the $ \gamma $- deformed set up in Section \ref{sec2}. In Section \ref{sec3}, we work out the detailed structure of the linearised perturbation equations. In Section \ref{sec4}, we compute the spin 2 spectrum taking specific examples of $ \gamma $- deformed ATD as well as NATD models. Finally, we draw our conclusion in Section \ref{sec5}, where we outline the possible structure of the spin 2 operator in the dual field theory.
\section{Deformed Gaiotto-Maldacena background in Type IIA}
\label{sec2}
To begin with, we review in detail the marginal deformations of 4d $ \mathcal{N}=2 $ SCFTs and their holographic duals in Type IIA. The marginal deformation of these theories leads towards a new class of $\mathcal{N}=1$ SCFTs in 4d whose holographic dual (in Type IIA) has been obtained by authors in \cite{Nunez:2019gbg}. These class of geometries are termed as $ \gamma $- deformed Gaiotto-Maldacena (GM) backgrounds those are constructed following an algorithm developed in \cite{Lunin:2005jy}-\cite{Gauntlett:2005jb}.

The seed solution is considered to be an eleven dimensional background in M theory. Typically, these class of geometries are obtained by applying an $ SL(3,R) $ transformation (along with a deformation parameter $ \gamma $) in M theory. Finally, the ten dimensional ($ \gamma $- deformed) Type IIA background is constructed following a circle ($ S^1 $) reduction\footnote{See \cite{Nunez:2019gbg} for details.} \cite{Nunez:2019gbg}
\begin{eqnarray} \label{dgm}
&d\hat s^2_{10} =   e^{- \frac{\hat \Phi}{2}} f_1 \Bigg[& ds_{AdS_5}^2   + \frac{f_2}{4f_1}  \Big(d\sigma^2 + d\eta^2\Big) + \frac{f_3}{4f_1} d\chi^2  + \frac{f_3 \sin^2\chi}{4f_1\big(1+ \gamma^2 f_3 f_4 \sin^2 \chi\big)} d\xi^2  \cr
&& +  \ \frac{f_4 }{4f_1\big(1+ \gamma^2 f_3 f_4  \sin^2 \chi\big)} \Big(d\beta - \gamma f_5 \sin\chi d\chi\Big)^2\Bigg] , 
 \end{eqnarray}
where $ \gamma $ is the deformation parameter such that the solution maps to the standard $\mathcal{N}=2$ supersymmetric Gaiotto-Maldacena background \cite{Reid-Edwards:2010vpm,Aharony:2012tz,Gaiotto:2009gz,Lozano:2016kum,Nunez:2019gbg} in the limit of the vanishing deformation, $ \gamma =0 $.

The $ AdS_5 $ line element could be expressed as
\begin{eqnarray} \label{ads5}
ds_{AdS_5}^2 =  -\cosh^2 r dt^2 + dr^2 + \sinh^2r d\Omega_3^2, 
 \end{eqnarray}
together with the metric functions $ f_i (\sigma , \eta) $ 
\begin{eqnarray} \label{f}
&&f_1 = \Bigg(\frac{2\dot{V}-\ddot{V}}{ V^{\prime \prime}}\Bigg)^{\frac{1}{2}}~;~ f_2 = f_1 \frac{2V^{\prime\prime}}{\dot{V}}~;~f_3 =  f_1 \frac{2V^{\prime\prime}\dot{V}}{\Delta}~;~ f_4 = f_1 \frac{4V^{\prime\prime}}{2\dot{V}-\ddot{V}} \sigma^2 \\
&&f_5 = 2 \Big(\frac{\dot{V}\dot{V}^\prime}{\Delta} - \eta\Big)~;~f_6 = \frac{2\dot{V}\dot{V}^\prime}{2\dot{V}-\ddot{V}}~;~f_7 =  - \frac{4 \dot{V}^2 V^{\prime\prime}}{\Delta}~;~f_8 = \Bigg[ (2)^{12}  \ \frac{4 \big(2\dot{V}-\ddot{V}\big)^3}{ V^{\prime\prime}\dot{V}^2 \Delta^2}\Bigg]^{\frac{1}{2}}.
 \end{eqnarray}

The dot and the prime of the potential function $ V(\sigma , \eta) $ are denoted as
\begin{eqnarray} 
&& \dot{V} = \sigma \partial_\sigma V \ ,  \  V^\prime = \partial_\eta V \ ,  \  \ddot{V} =  \sigma \partial_\sigma \dot{V} \ ,    \  V^{\prime\prime} = \partial_\eta^2 V \ ,   \  \dot{V}^\prime =  \sigma \partial_\sigma V^\prime \\
&&\Delta =  \big(2\dot{V} - \ddot{V} \big) V^{\prime\prime} + \big(\dot{V}^\prime\big)^2 . 
\end{eqnarray}

The background NS plus RR fluxes together with the dilaton could be expressed as \cite{Nunez:2019gbg}
\begin{eqnarray} 
\hat B_2 &=& \frac{1}{4}  \ \frac{1 }{1+ \gamma^2 f_3 f_4  \sin^2 \chi}  \bigg(f_5 d\Omega_2 (\chi, \xi) - \gamma f_3 f_4  \sin^2\chi d\xi \wedge d\beta\bigg),\\
\hat C_1 &=& \frac{1}{16}  \  \bigg(f_6 d\beta + \gamma \big(f_7 - f_5 f_6\big) \sin\chi d\chi\bigg),\\
\hat C_ 3 &=&  \frac{1}{64}  \ \frac{1}{1+ \gamma^2 f_3 f_4 \sin^2 \chi} f_7 d\beta \wedge d\Omega_2 (\chi, \xi),\\
e^{2 \hat \Phi} &=& \frac{ f_8}{  1+ \gamma^2 f_3 f_4 \sin^2 \chi}.
\label{dgmf}
 \end{eqnarray}
\section{Perturbations}
\label{sec3}
Given the above set up \eqref{dgm}-\eqref{dgmf}, we are now in a position to examine the spin-2 spectrum for the $\gamma$-deformed Gaiotto-Maldacena background those correspond to marginal deformations of $ \mathcal{N} =2$ SCFTs in four dimensions. 

Following the algorithm as in \cite{Itsios:2019yzp}, we express the $ \gamma $- deformed metric \eqref{dgm} (apart from the usual conformal factor) as sum of the AdS$_5$ factor and a five-dimensional internal space $\mathcal M_5^\gamma$ in the Einstein frame as 
\begin{eqnarray} \label{ref}
ds^2_{10} = ds_{AdS_5}^2 + ds_{\mathcal M_5^\gamma}^2 \ . 
\end{eqnarray}

We set the notation below, where we use the indices $M$ to denote the coordinates of the full ten-dimensional background. On the other hand, we use the indices $x^\mu$ to label the AdS$_5$ directions and $y^m$ for the $ \gamma $- deformed five-dimensional internal space $\mathcal M_5^\gamma$.

This results into a ten dimensional metric of the following form 
\begin{eqnarray} 
ds^2_{10} = \tilde g_{MN} dx^M dx^N =   \tilde g_{\mu \nu} (x) dx^\mu dx^\nu + \tilde g_{mn} (y) dy^m dy^n,
\end{eqnarray}
where one can split the full ten dimensional metric as
\begin{eqnarray}
  \tilde g_{MN} (x,y) =
	\begin{bmatrix} 
	 \tilde g_{\mu \nu} (x)  & 0  \\
	0 & \tilde g_{mn} (y)  \\
	\end{bmatrix}.
\end{eqnarray}

Next, we turn on the metric fluctuations along the AdS$_5$ 
\begin{eqnarray} 
\delta g_{\mu \nu} = e^{2A} h_{\mu \nu},
\end{eqnarray}
while keeping the other background fluctuations to zero.

This allows us to express the metric schematically as
\begin{eqnarray} \label{E}
ds_{E}^2 = e^{2A} \Bigg[\Big(\tilde g_{\mu \nu} (x) + h_{\mu \nu}\Big) dx^\mu dx^\nu +  \tilde g_{mn} (y) dy^m dy^n \Bigg].
\end{eqnarray}

A straightforward comparison between equations \eqref{E} and \eqref{dgm} yields
\begin{eqnarray} \label{A}
 A = - \frac{\hat \Phi(\eta, \sigma, \chi)}{4} + \frac{1}{2} \ln f_1(\eta, \sigma).
\end{eqnarray}

Next, we decompose the metric perturbation
\begin{eqnarray} 
 h_{\mu \nu} (x,y) =  \mathfrak{h}_{\mu \nu}^{[tt]} (x) F (y), 
\end{eqnarray}
along with the \emph{transverse} as well as the \emph{traceless} gauge $\tilde \nabla^\mu \mathfrak{h}_{\mu \nu}^{[tt]} (x) = 0 =  \tilde g^{\mu \nu} \mathfrak{h}_{\mu \nu}^{[tt]} (x)$.

Notice that, since the dilaton and the background fluxes do not change under metric perturbations, therefore the corresponding equations in \eqref{EOM} are trivially satisfied. On the other hand, the metric ($R_{MN}$) equation in \eqref{EOM} could be rephrased as
\begin{eqnarray} \label{RMN}
R_{MN} - \frac{1}{2} \partial_M \Phi  \partial_N \Phi - \frac{1}{2} \sum_{p=2}^{4} \gamma_p e^{\alpha_p \Phi} \Bigg[\Big(\mathcal A_p^2\Big)_{MN}- \beta_p g_{MN}  \mathcal A_p^2\Bigg]  = 0, 
\end{eqnarray}
where we denote the remaining fields collectively along with other notations
\begin{eqnarray} 
&&\mathcal A_p := \lbrace F_2, H_3, F_4 \rbrace \ ; \  \Big(\mathcal A_p^2\Big)_{MN} = \mathcal A_{M M_1 .. M_{p-1}}   \mathcal A_{N}^{M_1 .. M_{p-1}} \\
 &&\alpha_p := \Bigg(\frac{3}{2}, -1, \frac{1}{2}\Bigg) \ ;   \ \beta_p := \Bigg(\frac{1}{16}, \frac{1}{12},  \frac{3}{32} \Bigg) \ ;  \  \gamma_p := \Bigg(1, \frac{1}{2},  \frac{1}{6} \Bigg). 
\end{eqnarray}

Next, we use (\ref{Rh}) in order to simplify \eqref{RMN} while we simultaneously impose the following conditions on the background fields:\\
$ \bullet $ The only non zero part of $h_{MN}$ is $h_{\mu \nu}$ which means we set $h_{\mu m} = h_{m \mu} = h_{mn} = 0$.\\
$ \bullet $ $h_{\mu \nu}$ is transverse and traceless.\\
$ \bullet $ $A$ depends only on the internal coordinates $y^m$ namely, $A = A(y^m)$.\\
$ \bullet $ Finally, the collective background fields ($\tilde{\mathcal A}_p$) also depend on the internal coordinates $y^m$. 

Considering all the above points, one could finally express \eqref{RMN} as
\begin{eqnarray} \label{h3}
&& \Big(\tilde{g}^{\rho \sigma} \tilde{R}_{\sigma \mu} h_{\rho \nu} - \tilde{g}^{\kappa \sigma} \tilde{R}_{\nu \kappa \mu}^\rho h_{\sigma \rho}\Big) + \Big(\tilde{g}^{\rho \sigma} \tilde{R}_{\sigma \nu} h_{\rho \mu} - \tilde{g}^{\kappa \sigma} \tilde{R}_{\mu \kappa \nu}^\rho h_{\sigma \rho}\Big) 
-  \tilde{\nabla}^2 h_{\mu \nu} - 8 \tilde{\nabla}^{P} A \tilde{\nabla}_{P} h_{\mu \nu}  \cr
&&- 2h_{\mu \nu} \tilde{\nabla}^2 A - 16 h_{\mu \nu} \big( \tilde{\nabla} A\big)^2 + h_{\mu \nu}  \sum_{p=2}^{4} \beta_p \gamma_p e^{2(1-p) A +\alpha_p \Phi} \tilde {\mathcal A_p}^2 = 0.
\end{eqnarray}

Next, considering $AdS_5$ to be of unit radius and using the fact
\begin{eqnarray} \label {h4}
\tilde{g}^{\rho \sigma} \tilde{R}_{\sigma \mu} h_{\rho \nu} - \tilde{g}^{\kappa \sigma} \tilde{R}_{\nu \kappa \mu}^\rho h_{\sigma \rho} &=& -4h_{\mu \nu} - h_{\mu\nu} = -5h_{\mu\nu} \\
\tilde{g}^{\rho \sigma} \tilde{R}_{\sigma \nu} h_{\rho \mu} - \tilde{g}^{\kappa \sigma} \tilde{R}_{\mu \kappa \nu}^\rho h_{\sigma \rho} &=& -4h_{\nu \mu} - h_{\nu \mu} = -5h_{\mu\nu},
\end{eqnarray}
one could further simplify \eqref{h3} as
\begin{eqnarray} \label {h6}
\tilde{\nabla}^\alpha  \tilde{\nabla}_\alpha h_{\mu\nu} + \tilde{\nabla}^m  \tilde{\nabla}_m h_{\mu\nu} + 8 \tilde{\nabla}^{m} A \tilde{\nabla}_{m} h_{\mu \nu} + \mathcal{S}  h_{\mu\nu} = 0,
\end{eqnarray}
where we denote the above function as
\begin{eqnarray} \label{S}
 \mathcal{S}  = 10 + 2 \tilde{\nabla}^2  A
+ 16 \big(\tilde \nabla A\big)^2- \sum_{p=2}^{4} \beta_p \gamma_p e^{2(1-p) A +\alpha_p \Phi} \tilde {\mathcal A_p}^2.
\end{eqnarray}

The second and the third term of \eqref{h6} could be further simplified as 
\begin{eqnarray} \label {h7}
\tilde{\nabla}^m  \tilde{\nabla}_m h_{\mu\nu} + 8 \tilde{\nabla}^{m} A \tilde{\nabla}_{m} h_{\mu \nu} = e^{-8A} \tilde{\nabla}^m \Big[ e^{8A}  \tilde{\nabla}_m h_{\mu\nu} \Big] \equiv \mathcal L^{(1)} \Big[h_{\mu\nu} \Big],
\end{eqnarray}
which reveals \eqref{h6} in its final form
\begin{eqnarray} \label{h7}
 \tilde{\nabla}^\alpha  \tilde{\nabla}_\alpha h_{\mu\nu} + \mathcal L^{(1)} \Big[h_{\mu\nu} \Big]  
 + \mathcal{S} h_{\mu\nu} = 0. 
\end{eqnarray}

Here, the operator $\mathcal L^{(k)}$ is defined with respect to the internal coordinates ($ y^m $) as \cite{Passias:2018swc}
\begin{eqnarray} \label {LK}
 \mathcal L^{(k)} := e^{-8 kA} \tilde \nabla^m  \Big[ e^{8kA} \tilde \nabla_m  \Big].
\end{eqnarray}

For a massive graviton of mass $M$ propagating in $AdS_5$, one satisfies the {\it{Bachas Estes}} equation \cite{Itsios:2019yzp,Passias:2018swc,Bachas:2011xa,Passias:2016fkm,Richard:2014qsa,Gutperle:2018wuk,Chen:2019ydk,Speziali:2019uzn} of the form
\begin{eqnarray} \label{BE}
\Bigg[\tilde{\nabla}^\alpha  \tilde{\nabla}_\alpha + \Big(2 - M^2\Big)\Bigg] h_{\mu\nu}  = 0,
 \end{eqnarray}
which further simplifies \eqref{h7} by replacing AdS$_5$ derivatives ($ \tilde{\nabla}^\alpha  \tilde{\nabla}_\alpha $)
\begin{eqnarray} 
\label{3.18}
\Bigg[\mathcal L^{(1)} + \mathcal{S} + \Big(M^2 - 2 \Big)\Bigg] F(y) = 0,
 \end{eqnarray}
which is entirely defined with respect to internal directions ($ y^m $) where we define
\begin{eqnarray} \label{Lf}
 \mathcal L^{(1)} F =  \frac{1}{\sqrt{|\tilde {g}_{\mathcal {M}_5^\gamma}}|} \partial_m \Big(\sqrt{|\tilde {g}_{\mathcal {M}_5^\gamma}|}  \ \tilde{g}^{mn} \partial_n F \Big) + 8 \tilde{g}^{mn} \partial_m A \partial_n F, 
 \end{eqnarray}
where $\tilde {g}_{\mathcal {M}_5^\gamma}$ is the determinant of the $\gamma$-deformed five dimensional metric associated with the internal space \eqref{dgm}
\begin{eqnarray} 
\label{int}
d\hat s^2_{{\mathcal {M}_5^\gamma}} &=&   \frac{f_2}{4f_1}  \Big(d\sigma^2 + d\eta^2\Big) + \frac{f_3}{4f_1} d\chi^2  + \frac{f_3 \sin^2\chi}{4f_1\big(1+ \gamma^2 f_3 f_4 \sin^2 \chi\big)} d\xi^2  \nonumber\\
&& +  \ \frac{f_4 }{4f_1\big(1+ \gamma^2 f_3 f_4  \sin^2 \chi\big)} \Big(d\beta - \gamma f_5 \sin\chi d\chi\Big)^2.
 \end{eqnarray}

Using (\ref{int}) together with (\ref{A}) one can finally express \eqref{Lf} as
\begin{eqnarray} \label{LF}
 \mathcal L^{(1)} F &=& \frac{2}{f_2} \Bigg[5\Big(\partial_\eta f_1 \partial_\eta F + \partial_\sigma f_1 \partial_\sigma F\Big) + \frac{1}{f_3 f_4 (\gamma^2 f_3 f_4 + \csc^2\chi)} f_1 \Big(2 f_2 \Big(\csc^2\chi f_3  \partial_\beta^2 F \cr
 && \gamma^4 f_3^2 f_4^3  \partial_\xi^2 F + f_4 \Big(\csc^2\chi \Big(\Big(\cot \chi - 2 \partial_\chi \hat\Phi\Big) \partial_\chi F +  \partial_\chi^2 F +  \csc^2\chi \partial_\xi^2 F\Big) + \cr 
 && 2\gamma^2 f_3^2  \partial_\beta^2 F + \gamma^2 f_5^2 \partial_\beta^2 F + 2 \gamma \csc \chi f_5 \Big(\Big(\cot\chi - \partial_\chi \hat\Phi\Big)\partial_\beta F + \partial_\chi \partial_\beta F\Big)\Big) + \cr 
 &&\gamma^2 f_3 f_4^2 \Big(-\Big(2\partial_\chi \hat\Phi + \cot\chi\Big)\partial_\chi F + \partial_\chi^2 F + 2\csc^2 \chi \partial_\xi^2 F + \gamma^2 \sin^2\chi f_3^2  \partial_\beta^2 F + \cr
 && \gamma^2  \sin^2\chi f_5^2  \partial_\beta^2 F + 2 \gamma \sin \chi f_5 \Big( \partial_\chi \partial_\beta F - \partial_\chi \hat\Phi \partial_\beta F\Big)\Big)\Big) + \csc^2\chi \Big(2f_4 \Big(\partial_\eta f_3  \cr 
 &&\partial_\eta F + \partial_\sigma f_3 \partial_\sigma F + f_3 \Big(-2 \partial_\eta \hat \Phi  \partial_\eta F - 2 \partial_\sigma \hat\Phi  \partial_\sigma F + \partial_\eta^2 F +  \partial_\sigma^2 F\Big)\Big) \cr
 &&+ f_3 \Big(\partial_\eta f_4  \partial_\eta F + \partial_\sigma f_4  \partial_\sigma F\Big)\Big) - \gamma^2 f_3^2 f_4  \Big(\partial_\eta f_4  \partial_\eta F + \partial_\sigma f_4  \partial_\sigma F + f_4 \Big(4\partial_\eta \hat\Phi \cr
 &&\partial_\eta F - 2 \Big(-2 \partial_\sigma \hat\Phi \partial_\sigma F + \partial_\eta^2 F +  \partial_\sigma^2 F \Big)\Big)\Big) \Big) \Bigg].  
 \end{eqnarray}
It is quite trivial to notice that, in the limit $\gamma \rightarrow 0$, together with proper redefinition of $f_i$'s, the expression \eqref{LF} boils down into the original Gaiotto-Maldacena background \cite{Itsios:2019yzp}.
For the generic class of Gaiotto-Maldacena geometries characterised by the functions $f_i (\sigma, \eta)$ the universal solution of \eqref{3.18} is discussed in \cite{Chen:2019ydk}.
However, in the presence of $\gamma$-deformation the universal solution turns out to be highly non-trivial. 
Therefore, in present paper we concern two specific examples namely $\gamma$-deformed Abelian and non-Abelian T-dual backgrounds. 
\section{Spin $ 2 $ spectrum}
\label{sec4}
Below, we discuss in detail the spectrum associated with massive graviton fluctuations taking two specific examples within the deformed Gaiotto-Maldacena class of geometries. These are the $ \gamma $- deformed Abelian T dual (ATD) and non-Abelian T dual (NATD) solutions those were obtained by authors in \cite{Nunez:2019gbg}. The addition of flavour branes complicates the scenario and the corresponding equations of motion are hardly tractable analytically. We outline these issues in the Appendix \ref{appen c}.
\subsection{$ \gamma $- deformed ATD}
The potential function for the ATD case takes the form \cite{Lozano:2016kum}
\begin{eqnarray} \label{V atd}
V_{\text{ATD}} (\sigma,\eta) = \ln \sigma - \frac{1}{2} \sigma^2 + \eta^2 \ . 
 \end{eqnarray}

The associated functions $f_i(\sigma,\eta)$ in \eqref{f} turn out to be \footnote{There is a shift symmetry in the expression \eqref{V atd} namely $V_{\text{ATD}} (\sigma,\eta) \rightarrow V_{\text{ATD}} (\sigma,\eta) + A \eta$ ; where $A=$ const.  This shift symmetry can be interpreted as a residual diffeomorphism/ gauge that preserves the metric as well as the background fluxes. By using this symmetry, one can therefore gauge away $f_5$ as $\tilde{f}_5 (= f_5+ 2\eta)$, such that $\tilde{f}_5$ vanishes for the potential function \eqref{V atd}. This restores the $\psi$-isometry for the ($ \gamma $- deformed) ATD solution. This gauge redundancy is absent for the non-Abelian T-dual solution.}
\begin{eqnarray} \label{f-atd}
&&f_1 =1~;~ f_2 = \frac{4}{1-\sigma^2}~;~f_3 =  1-\sigma^2~;~ f_4 =4 \sigma^2 \\
&&f_5 = - 2\eta ~;~f_6 =0~;~f_7 =  - 2 \big(1-\sigma^2\big)^2~;~f_8 = \frac{64}{1-\sigma^2}. 
 \end{eqnarray}

Using \eqref{V atd}, the expression in \eqref{LF} takes the form 
\begin{eqnarray} \label{LFATD}
 \mathcal L^{(1)} F &=& \frac{1}{\sigma^2 \big(\sigma^2 -1\big)} \Bigg[\Big(-4 \gamma^2 \sigma^2\Big(4\eta^2 + \Big(\sigma^2 -1\Big)^2\Big) \sin^2\chi + \sigma^2 -1\Big) \partial_\beta^2 F \cr
 && \sigma \Big(4\sigma \Big(\Big(4\gamma^2 \sigma^2\Big(\sigma^2-1\Big) - \csc^2\chi\Big)\partial_\xi^2 F + 4 \gamma \eta \sin\chi \partial_\chi \partial_\beta F - \partial_\chi^2 F \cr
 && - \cot\chi \partial_\chi F\Big) - \sigma \Big(\sigma^2 -1\Big)^2 \partial_\eta^2 F -  \sigma \Big(\sigma^2 -1\Big)^2  \partial_\sigma^2 F + \Big(-5 \sigma^4 + 6\sigma^2 -1\Big) \cr
 &&\partial_\sigma F\Big) + 16 \gamma \eta \sigma^2 \cos\chi \partial_\beta F \Bigg]. 
\end{eqnarray}

Using the change of variables
\begin{eqnarray} 
\eta = 2 \psi \ ;  \  \sigma = \sin\alpha
 \end{eqnarray}
and expanding the L.H.S. of \eqref{3.18} upto $ \mathcal{O}(\gamma) $ we find
\begin{eqnarray} \label{eq}
&&\partial_\alpha^2 F + \Big(\cot\alpha - 3\tan\alpha\Big) \partial_\alpha F  + \frac{1}{\sin^2\alpha} \partial_\beta^2 F + \frac{\cos^2\alpha}{4} \partial_\psi^2 F + \frac{4}{\cos^2\alpha} \nabla_{(2)}^2 (\chi,\xi)F \cr
&& - 32 \frac{\psi \gamma}{\cos^2\alpha} \sin\chi \Big(\partial_\chi \partial_\beta F + \cot\chi \partial_\beta F\Big)+ M^2 F = 0,
 \end{eqnarray}
where $ \nabla_{(2)}^2 (\chi,\xi) = \partial_\chi^2 + \cot\chi \partial_\chi + \csc^2\chi \partial_\xi^2$, is the Laplace operator corresponding to $S^2 (\chi, \xi)$. 

Next, we propose an ansatertz of the form
\begin{eqnarray} \label{ans}
F = e^{i m \beta} e^{i q \xi} \ \tilde{F}(\alpha , \chi , \psi),
 \end{eqnarray}
where $ \beta $ and $ \xi $ are the isometry directions of the internal manifold and $\lbrace m, q\rbrace $ are respectively the associated quantum numbers. 

This further simplifies \eqref{eq} as
\begin{eqnarray} \label{e1}
&&\partial_\alpha^2 \tilde{F} + \Big(\cot\alpha - 3\tan\alpha\Big) \partial_\alpha \tilde{F}  - \frac{1}{\sin^2\alpha} m^2 \tilde{F} + \frac{\cos^2\alpha}{4} \partial_\psi^2 \tilde{F} + \frac{4}{\cos^2\alpha}  \cr
&& \Big(\partial_\chi^2 \tilde{F} + \cot\chi \partial_\chi \tilde{F} - q^2 \csc^2\chi \tilde{F} \Big) + M^2 \tilde{F}  \nonumber\\ &&- 32 (im) \frac{\psi \gamma}{\cos^2\alpha} \sin\chi \Big(\partial_\chi  \tilde{F} + \cot\chi \tilde{ F}\Big) 
= 0. 
 \end{eqnarray}
Interestingly, for $m=0$ mode the imaginary part in \eqref{e1} identically vanishes and we are left only with the real part.
On the other hand, for $m\neq0$ mode, the solution corresponding to the imaginary part ($\tilde{F}_{\text{imaginary}} \sim \csc\chi$) yields a diverging contribution near $\chi = \{0,\pi\}$. As a result, the corresponding graviton fluctuation \eqref{ans} blows up. Therefore, these nonzero ($ m \neq 0 $) modes are discarded for the purpose of the present analysis.
\subsection{$m=0$ mode}
As mentioned previously, for $m=0$ mode, the imaginary part of \eqref{e1} vanishes and the equation for $\tilde{F}$ takes the following form
\begin{eqnarray} \label{new}
&&\partial_\alpha^2 \tilde{F} + \Big(\cot\alpha - 3\tan\alpha\Big) \partial_\alpha \tilde{F}   + \frac{\cos^2\alpha}{4} \partial_\psi^2 \tilde{F} + \frac{4}{\cos^2\alpha}  \cr
&& \Big(\partial_\chi^2 \tilde{F} + \cot\chi \partial_\chi \tilde{F} - q^2 \csc^2\chi \tilde{F} \Big) 
+ M^2 \tilde{F} = 0. 
 \end{eqnarray}
 In order to proceed further, we propose a separation of variable of the form
\begin{eqnarray} \label{sep}
\tilde{F} (\alpha , \chi, \psi) = F_\alpha (\alpha) F_\psi (\psi) F_\chi (\chi). 
 \end{eqnarray}
Plugging \eqref{sep} in \eqref{new} we obtain 
\begin{eqnarray} \label{fi1}
&&\partial_\alpha^2  F_\alpha (\alpha) + \Big(\cot\alpha - 3\tan\alpha\Big) \partial_\alpha  F_\alpha (\alpha)  + \Bigg[ M^2  +  \frac{\cos^2\alpha}{4} \frac{1}{F_\psi(\psi)}  \partial_\psi^2 F_\psi (\psi) \cr
&&+  \frac{4}{\cos^2\alpha} \frac{1}{F_\chi (\chi)} \Big(\partial_\chi^2 F_\chi (\chi) + \cot\chi \partial_\chi F_\chi (\chi)  - q^2 \csc^2\chi F_\chi (\chi) \Big) \Bigg] F_\alpha (\alpha) = 0, 
 \end{eqnarray}
where the $ \psi $- equation has a solution of the form
\begin{eqnarray} \label{psi1}
 F_\psi (\psi)  = A_1 \sin(2n\psi) + A_2 \cos(2n\psi).
\end{eqnarray}
Here, $\psi$ takes values within the interval $[0, \frac{\pi}{n}]$ where $ n $ is an integer. As a result, the corresponding function $F_\psi (\psi)$ is also periodic in the given interval. 

Similarly the equation for $F_\chi(\chi)$ in \eqref{fi1} takes the form 
\begin{eqnarray} \label{chi1}
\partial_\chi^2 F_\chi (\chi) + \cot\chi \partial_\chi F_\chi (\chi) + \Big(K^2 - q^2 \csc^2\chi\Big) F_\chi (\chi)  = 0, 
 \end{eqnarray}
and yields a solution in terms of the Legendre functions  $P , Q$ \cite{Abramowitz}
\begin{eqnarray} \label{chi new}
 F_\chi (\chi) = B_1 P \Bigg[m , q , \cos\chi\Bigg] + B_2 Q \Bigg[ n , q , \cos\chi\Bigg],
 \end{eqnarray}
where $m =\frac{1}{2} \Big(\sqrt{4K^2+1} - 1\Big)$ and $ n= \frac{1}{2} \Big(\sqrt{4K^2+1} - 1\Big)$ are integers such that $F_\chi (\chi)$ is smooth within the interval $0 \leq \chi \leq \pi$, together with $B_{1,2}$ as constants.

Finally, by plugging \eqref{psi1}-\eqref{chi new} into \eqref{fi1} we find
\begin{eqnarray} 
\partial_\alpha^2  F_\alpha (\alpha) + \Big(\cot\alpha - 3\tan\alpha\Big) \partial_\alpha  F_\alpha (\alpha)
+ \Bigg[M^2 - \frac{4K^2}{\cos^2\alpha} - n^2 \cos^2\alpha\Bigg] F_\alpha (\alpha)   
= 0.
 \end{eqnarray}
Next, we implement the change in variable $z = \sin^2\alpha$ which finally yields
\begin{eqnarray} \label{new z}
z\big(1-z\big) \partial_z^2  F_z (z) + \big(1-3z\big) \partial_z  F_z(z) 
+ \Bigg[\frac{M^2}{4} - \frac{K^2}{1-z} - \frac{n^2 \big(1-z\big)}{4}\Bigg] F_z (z)   
= 0.  
 \end{eqnarray}
Notice that, the above equation clearly differs from \cite{Itsios:2019yzp} as we do not have the contribution ($ \sim \ell (\ell +1) $ ) due to the $ S^2 (\chi , \xi)$ as well as contribution ($ m $) due to $ S^1 (\beta)$ of the internal manifold. 
This stems from the fact that the original $ SU(2) $ symmetries of $ S^2 (\chi , \xi) $ are broken down to a single $ U(1) $ isometry (corresponding to the $ \xi $ direction (see \eqref{int}) in the presence of the $ \gamma $- deformation and we fixed $m=0$ in our analysis. 
\subsection{Zero modes: $n=0$} 
Setting $n=0$ for zero modes, one finds
\begin{eqnarray} \label{new n0}
z\big(1-z\big) \partial_z^2  F_z (z) + \big(1-3z\big) \partial_z  F_z(z) 
+ \Bigg[\frac{M^2}{4} - \frac{K^2}{1-z} \Bigg] F_z (z)   
= 0.  
 \end{eqnarray}
A particular solution of \eqref{new n0} exists only for $K = 0$ mode and in the small $z \sim 0$ limit which is equivalent to the $ \sigma \sim 0 $ limit. 
A closer look reveals that the corresponding solution can be expressed in terms of the Bessel function and Neumann function of the form
\begin{eqnarray} \label{z n}
F(z)_{z \sim 0} =   B_3 J_0 \big(M\sqrt{z}\big) + B_4 Y_{0} \big(M\sqrt{z}\big), 
 \end{eqnarray}
where $M$(mass of the graviton)$ > 0$ and $B_{3,4}$ are constants. 
On the other hand, the function $Y_{0} \big(M\sqrt{z}\big)$ diverges like $\log z$ close to $z \sim 0$, therefore we set the constant, $B_4 = 0$ in \eqref{z n}. 

Combining all these facts together, the final solution takes the form 
\begin{eqnarray} \label{z n f}
F(z)_{z \sim 0} =   B_3 J_0 \big(M\sqrt{z}\big) , 
 \end{eqnarray}
where $M > 0$. 

The above analysis shows that $F_z(z)$ in \eqref{z n f} is indeed a smooth function in the small $ z $ limit without imposing any further constraints. Contrary to the undeformed case \cite{Itsios:2019yzp}, here we obtain a continuous spectrum for the massive graviton. As mentioned previously, this is an artefact of the absence of the spherical symmetry in the presence of the $\gamma$- deformation. 
\subsubsection{$ n\neq 0 $ case}
The $ n\neq 0 $ case can be dealt through the WKB method of \cite{Minahan:1998tm}-\cite{Russo:1998by} where the graviton mass ($M$) is considered to be very large. 

Introducing the following change in coordinates
\begin{eqnarray} \label{r}
r= \frac{z}{1-z}~;~ r \in [0,+\infty) 
 \end{eqnarray}
the graviton equation \eqref{new z} could be rephrased as\footnote{Here we replace $ F_z (z)$ by $ \Psi(r) $ in the transformed coordinates.}
\begin{eqnarray} \label{wkb}
\partial_r \Bigg[p(r) \partial_r \Psi(r)\Bigg] + \Bigg[M^2 w(r) + q(r)\Bigg] \Psi(r) = 0, 
 \end{eqnarray}
where we define the above functions as
\begin{eqnarray} \label{pwq}
p(r) =  \frac{r}{1+r} \ ;  \ w(r) =  \frac{1}{4\big(1+r\big)^3} \ ; \  q(r) = -  \Bigg[\frac{K^2}{\big(1+r\big)^2} + \frac{n^2}{\big(1+r\big)^4}\Bigg]. 
 \end{eqnarray}

Expanding the above functions \eqref{pwq} near $r \sim 0$ one finds 
\begin{eqnarray} \label{pwq 0}
p(r) &\approx& r + \mathcal{O}(r^2), \\
w(r) &\approx& \frac{1}{4} + \mathcal{O}(r), \\
q(r) &\approx& - \Big(K^2 + \frac{n^2}{4}\Big) + \Big(2K^2 + n^2\Big)r + \mathcal{O}(r^2),
 \end{eqnarray}
which can be further compared to yield the associated exponents as
\begin{eqnarray} \label{r=0}
p(r) &\approx& p_1 r^{s_1} \ ;  \ p_1 = s_1 = 1 ,                  \\
w(r) &\approx& w_1 r^{s_2} \ ; \ w_1 =  \frac{1}{4}  \ ,  \ s_2 = 0,   \\
q(r) &\approx&q_1 r^{s_3} \ ;  \ q_1 = - \Big(K^2 + \frac{n^2}{4}\Big)  \ ,  \ s_3 = 0. 
\label{4.26}
 \end{eqnarray}

On a similar note, an expansion near $r \rightarrow \infty$ reveals
\begin{eqnarray} \label{pwq inf}
p(r) &\approx& 1 + \mathcal{O}(r^{-1}), \\
w(r) &\approx& \frac{1}{4r^3} + \mathcal{O}(r^{-4}), \\
q(r) &\approx& - \frac{K^2}{r^2} + \frac{2K^2}{r^3} - \frac{1}{r^4} \Big(3K^2 - \frac{n^2}{4}\Big)+  \mathcal{O}(r^{-5}),
\label{4.32}
 \end{eqnarray}
which can be further compared to decode the associated exponents as 
\begin{eqnarray} \label{r=inf}
p(r) &\approx& p_2 r^{t_1} \ ;  \ p_2 = 1 \ ,  \  t_1 = 0 \ ,                  \\
w(r) &\approx& w_2 r^{t_2} \ ; \ w_2 =  \frac{1}{4}  \ ,  \ t_2 = -3  \ ,   \\
q(r) &\approx&q_2 r^{t_3} \ ;  \ q_2 = - K^2  \ ,  \ t_3 = -2. 
 \end{eqnarray}

Finally, we use the above information in order to express the graviton mass \cite{Minahan:1998tm}-\cite{Russo:1998by}
\begin{eqnarray} \label{mass}
M^2 = \frac{\pi^2}{\zeta^2} \lambda \Bigg[\lambda - 1 + \frac{\alpha_2}{\alpha_1} +   \frac{\beta_2}{\beta_1} \Bigg],   
\end{eqnarray}
where $ \lambda (\geq 1) $ is a positive continuous parameter.

Here, we define the function $ \zeta $ through the following integral
\begin{eqnarray} 
\zeta = \int_{0}^{\infty} dr \sqrt{\frac{w}{p}} = \frac{1}{2} \int_{0}^{\infty} dr  \frac{1}{\sqrt{r} (1+r)} = \frac{\pi}{2}, 
 \end{eqnarray}
together with the other parameters as 
\begin{eqnarray} \label{ref}
\alpha_1 &=& s_2 - s_1 + 2 = 1,  \cr
\alpha_2 &=& |s_1 - 1| = 0, \cr
\beta_1 &=& t_1 - t_2 - 2 = 1,  \cr
\beta_2 &=& \sqrt{\big(t_1 -1\big)^2 - 4 \frac{q_2}{p_2}}  = \sqrt{1+4K^2}.
 \end{eqnarray}

Plugging \eqref{ref} into \eqref{mass} we finally obtain 
\begin{eqnarray} \label{mass sep}
M^2 = 4 \lambda \Big(\lambda - 1 + \sqrt{1+4K^2} \Big). 
 \end{eqnarray}
 
Here, couple of points are to be noticed. First of all, the angular momentum quantum number ($ \sim \ell $) does not appear in \eqref{mass sep} since the symmetries of the two sphere $ S^2(\chi, \xi) $ is broken (as a result of the $ \gamma $- deformation) therefore it is not a conserved quantity anymore. Secondly, we notice that the spectrum does not contain any information about the quantum number $ n $ as also shown previously by the authors in \cite{Itsios:2019yzp}. This is due to the fact that the $ n^2 $ term is suppressed in the expression \eqref{ref} therefore it does not show up in \eqref{mass sep}.
\subsection{$ \gamma $- deformed NATD}
Technically, the analysis for the $ \gamma $- deformed NATD is quite similar in spirit to that with the $ \gamma $- deformed ATD example. The corresponding potential function is given by \cite{Lozano:2016kum}
\begin{eqnarray} \label{V natd}
V_{\text{NATD}} (\sigma,\eta) = \eta\Big(\ln \sigma - \frac{1}{2} \sigma^2\Big) + \frac{1}{3} \eta^3, 
 \end{eqnarray}
together with the associated metric functions  $f_i(\sigma,\eta)$
\begin{eqnarray} \label{f-natd}
&&f_1 =1~;~ f_2 = \frac{4}{1-\sigma^2}~;~f_3 =  \frac{4\eta^2 (1-\sigma^2)}{4\eta^2 + (1-\sigma^2)^2}~;~ f_4 =4 \sigma^2 ,\\
&&f_5 = - \frac{8\eta^3}{4\eta^2 + (1-\sigma^2)^2} ~;~f_6 =\big(1-\sigma^2\big)^2~;~f_7 =  - \frac{8\eta^3 (1-\sigma^2)^2}{4\eta^2 + (1-\sigma^2)^2}, \\
&&f_8 = \frac{256}{(1-\sigma^2)(4\eta^2 + (1-\sigma^2)^2)}. 
 \end{eqnarray}

Then corresponding expression in \eqref{LF} takes the form 
\begin{eqnarray} \label{LFNATD}
 \mathcal L^{(1)} F &=& - \frac{1}{\eta^2 \sigma^2 (\sigma^2 -1)} \Bigg[\eta^2 \Big(16 \gamma^2 \eta^2 \sigma^2 \sin^2\chi - \sigma^2 +1\Big) \partial_\beta^2 F - \sigma^2 \csc^2\chi \cr
 && \Big(16 \gamma^2 \eta^2 \sigma^2 \Big(\sigma^2 -1 \Big) \sin^2\chi - 4\eta^2 - \Big(\sigma^2 -1 \Big)^2\Big) \partial_\xi^2 F - 16 \gamma \eta^3 \sigma^2 \sin\chi  \partial_\chi \partial_\beta F \cr
 && -16 \gamma \eta^3 \sigma^2 \cos\chi \partial_\beta F + \eta^2 \sigma^6 \partial_\eta^2 F + 5 \eta^2 \sigma^5 \partial_\sigma F - 2 \eta^2 \sigma^4 \partial_\eta^2 F + 6 \eta^2 \sigma^3 \cr
 && \partial_\sigma F + 4 \eta^2 \sigma^2 \partial_\chi^2 F + \eta^2 \sigma^2 \partial_\eta^2 F + \eta^2 \Big(\sigma^3 -\sigma\Big)^2 \partial_\sigma^2 F + \sigma^2 \Big(4\eta^2 + \sigma^4 - 2 \sigma^2 +1\Big) \cr
 && \cot\chi \partial_\chi F + \eta^2 \sigma \partial_\sigma F + \sigma^6 \partial_\chi^2 F + 2 \eta \sigma^6 \partial_\eta F - 2 \sigma^4 \partial_\chi^2 F - 4 \eta \sigma^4 \partial_\eta F \cr
 &&+ \sigma^2 \partial_\chi^2 F + 2 \eta \sigma^2 \partial_\eta F\Bigg]. 
   \end{eqnarray}

Using the following change in variables 
\begin{eqnarray} 
\eta = 2 \psi \ ;  \  \sigma = \sin\alpha , 
 \end{eqnarray}
and retaining terms upto leading order in the deformation parameter one finds
\begin{eqnarray} \label{FNATD}
 &&\partial_\alpha^2 F  + \Big(\cot\alpha - 3\tan\alpha\Big) \partial_\alpha F  + \frac{1}{\sin^2\alpha} \partial_\beta^2 F  \cr
&&+ \frac{\cos^2\alpha}{4} \bigg( \partial_\psi^2 F + \frac{2}{\psi} \partial_\psi F + \frac{1}{\psi^2}  \nabla_{(2)}^2 (\chi,\xi)F\bigg) + \frac{4}{\cos^2\alpha} \nabla_{(2)}^2 (\chi,\xi)F \cr
&&- 32 \frac{\psi \gamma}{\cos^2\alpha} \sin\chi \Big(\partial_\chi \partial_\beta F + \cot\chi \partial_\beta F\Big) + M^2 F = 0. 
 \end{eqnarray}
 
Like before, we use an ansatz of the following form
\begin{eqnarray} \label{ansn}
F = e^{i m \beta} e^{i q \xi} \ \tilde{F}(\alpha , \chi , \psi),
\end{eqnarray}
which splits the equation in \eqref{FNATD} into the form 
\begin{eqnarray} \label{e1n}
&&\partial_\alpha^2 \tilde{F} + \Big(\cot\alpha - 3\tan\alpha\Big) \partial_\alpha \tilde{F}  - \frac{1}{\sin^2\alpha} m^2 \tilde{F} + \frac{\cos^2\alpha}{4} \cr
&&\Bigg[\partial_\psi^2 \tilde{F} + \frac{2}{\psi} \partial_\psi \tilde{F} + \frac{1}{\psi^2}  \Big(\partial_\chi^2 \tilde{F} + \cot\chi \partial_\chi \tilde{F} - q^2 \csc^2\chi \tilde{F} \Big)\Bigg] +  \cr
&&\frac{4}{\cos^2\alpha}  \Big(\partial_\chi^2 \tilde{F} + \cot\chi \partial_\chi \tilde{F} - q^2 \csc^2\chi \tilde{F} \Big)\nonumber\\&&  - 32 (im) \frac{\psi \gamma}{\cos^2\alpha} \sin\chi \Big(\partial_\chi  \tilde{F} + \cot\chi \tilde{ F}\Big) +  M^2 \tilde{F} = 0. 
 \end{eqnarray}

The next step would be to use the following ansatz
\begin{eqnarray} \label{sep n}
\tilde{F}(\alpha , \chi , \psi) =  F_\alpha(\alpha) F_\chi(\chi) F_\psi(\psi), 
 \end{eqnarray}
and thereby considering only on $m=0$ mode we obtain
\begin{eqnarray} \label{Fnn} 
&&\partial_\alpha^2  F_\alpha (\alpha) + \Big(\cot\alpha - 3\tan\alpha\Big) \partial_\alpha  F_\alpha (\alpha)  + \Bigg[ M^2 +  \frac{\cos^2\alpha}{4} \frac{1}{F_\psi(\psi)} \bigg(\partial_\psi^2 F_\psi (\psi) \cr
&&+ \frac{2}{\psi} \partial_\psi F_\psi (\psi) + \frac{1}{\psi^2} \frac{1}{F_\chi (\chi)} \Big(\partial_\chi^2 F_\chi (\chi) + \cot\chi \partial_\chi F_\chi (\chi)  - q^2 \csc^2\chi F_\chi (\chi) \Big)  \bigg) \cr
&&+  \frac{4}{\cos^2\alpha}   \frac{1}{F_\chi (\chi)} \Big(\partial_\chi^2 F_\chi (\chi) + \cot\chi \partial_\chi F_\chi (\chi)  - q^2 \csc^2\chi F_\chi (\chi) \Big) \Bigg]   F_\alpha (\alpha) = 0 , 
 \end{eqnarray}
Like before, we obtain the same solution for the $F_\chi (\chi)$ as in the Abelian T-dual case in \eqref{chi new}. 
However, in this case
solution for the $ \psi $- equation takes the form 
\begin{eqnarray} \label{rho}
F_\psi (\psi) =  \frac{e^{-2 n \psi}}{\psi} - \frac{K^2}{4n^2 \psi} . 
 \end{eqnarray}
 
Finally, using \eqref{rho} and introducing the new coordinate $ z= \sin^2 \alpha $, one arrives at an equation 
\begin{eqnarray} \label{zan}
z\big(1-z\big) \partial_z^2  F_z (z) + \big(1-3z\big) \partial_z  F_z(z) 
+ \Bigg[\frac{M^2}{4} - \frac{K^2}{1-z} + \frac{n^2 \big(1-z\big)}{4}\Bigg] F_z (z)   
= 0,  
 \end{eqnarray}
which is almost identical to that of the $ \gamma $- deformed ATD \eqref{new z} except for the sign in front of the $ n^2 $ term. Notice that, the case with the zero ($ K=0 $ and $n=0$) modes is exactly identical to what we have found for the deformed ATD case. On the other hand, for the non zero modes, one can show that the $ n^2 $ term is largely suppressed both in the $ r \sim 0 $ and $ r \rightarrow \infty $ limits which reveals an identical spectra as in the deformed ATD case. 
\section{Concluding remarks}
\label{sec5}
Before we conclude, a couple of remarks are in order. First of all, it is worthwhile to mention that the graviton spectrum does not receive any correction due to $ \mathcal{S}  =  2 + \mathcal{O}(\gamma^2) $ at leading order in the deformation (see \eqref{3.18}). Therefore, at leading order, we have a nontrivial cancellation $ \mathcal{S}  -  2 \sim 0$ like in the undeformed case \cite{Itsios:2019yzp}. This plays a crucial role while solving the graviton spectrum at leading order in the $ \gamma $- deformation.
In the literature we analytically solve the graviton spectrum in the presence of $\gamma$-deformation under some specific choice of the quantum number $m = 0$. 

Finally, we would like comment on the dual spin 2 operator associated to the $ \mathcal{N}=1 $ superconformal quiver. As mentioned previously, the $ SU(2)_{\chi, \xi} $ charges of the original Gaiotto-Maldacena background are broken down to a single $ U(1)_{\xi} $ charge ($ q $) as a result of the $ \gamma $- deformation. The remaining $ U(1)_{\beta} $ charge ($ m $) is set to be zero.

Notice that, the spin 2 operator that we are going to propose is associated to a dual superconformal quiver that is \emph{unbounded} unless we add flavour degrees of freedom. The spin 2 operator has a continuous spectrum whose dimension could be typically expressed in the form, $ \Delta = 4f(\lambda, K)+ q $, where $ \lambda $ is some continuous parameter. 

The dual spin 2 operator can be constructed taking scalars ($ \phi^k $) from tensor multiplets (those also connect different vector multiplets and thereby carry color indices ($ k $)) and the gauge fields ($ A^I $) of the vector multiplet those are sourced due to the color D4 branes.  The resulting operator could be schematically expressed as
\begin{eqnarray}
\mathcal{O}_{\mu \nu} \sim \text{Tr}(\phi^{k_i}A^{I_i})T_{\mu \nu}~;~\sum_i k_i +\sum_i I_i =q
\end{eqnarray}
where the operator carries a $R$-charge of the form $ (q,m=0) $. However, the explicit expression of these dual operators is left for future investigations.

It would be indeed an interesting project to repeat the above analysis in the presence of flavours and construct the dual operator spectrum using a holographic set up. However, as we outline in the Appendix \ref{appen c}, the corresponding massive graviton equations are quite involved and therefore special numeric techniques must be adopted to solve the spectrum.
\acknowledgments
Its a pleasure to thank G. Itsios for clarifying some issues. One of the authors DR is indebted to the authorities of IIT Roorkee for their unconditional support towards researches in basic sciences. DR also acknowledges The Royal Society, UK for financial support.
\appendix
\section{Type-IIA supergravity equations}
In this Appendix, we briefly discuss the Type-IIA supergravity solutions of \cite{Passias:2018swc}. The field strengths of massive Type-IIA supergravity can be expressed as
\begin{eqnarray}
H_3 = dB_2 \ ; \ F_2 = dC_1 + F_0 B_2 \ ; \ F_4 = dC_3 - H_3 \wedge C_1 + \frac{1}{2} F_0 B_2 \wedge B_2, 
\end{eqnarray}
together with the Bianchi identities which take the following form
\begin{eqnarray}
d H_3 =0 \ ; \ dF_2 = F_0 H_3  \ ; \ d  F_4= H_3 \wedge F_2.
\end{eqnarray}

The corresponding equations of motion are given by  
\begin{eqnarray} \label{EOM}
0 &=& R_{MN} - \frac{1}{2} \partial_M \Phi \partial_N \Phi -\frac{1}{16} F_0^2 e^{\frac{5\Phi}{2}} g_{MN} - \frac{1}{2} e^{\frac{3\Phi}{2}} \Bigg(F_{MP} F_{N}^P -  \frac{1}{16} g_{MN} \big(F_2\big)^2\Bigg) \cr
&& - \frac{1}{12} e^{\frac{\Phi}{2}} \Bigg(F_{MPQR} F_{N}^{PQR} - \frac{3}{32} g_{MN} \big(F_4\big)^2\Bigg) - \frac{1}{4} e^{-\Phi} \Bigg(H_{MPQ} H_{N}^{PQ} - \frac{1}{12} g_{MN} \big(H_3\big)^2\Bigg) \ ,  \cr
0 &=& \nabla^M \nabla_M \Phi - \frac{5}{4} F_0^2 e^{\frac{5\Phi}{2}} - \frac{3}{8} e^{\frac{3\Phi}{2}} \big(F_2\big)^2 - \frac{1}{96} e^{\frac{\Phi}{2}} \big(F_4\big)^2 + \frac{1}{12} e^{-\Phi} \big(H_3\big)^2 \ , \cr
0 &=& \nabla^M \big(e^{-\Phi} H_{MNP}) - F_0 e^{\frac{3\Phi}{2}} F_{NP} - \frac{1}{2} e^{\frac{\Phi}{2}} F_{NPQR}F^{QR} + \frac{1}{2. 4 ! 4 !} \epsilon_{M_1 ... M_8 N P} F^{M_1 ... M_4} F^{M_5 ... M_8} \ ,  \cr
0 &=& \nabla^M \big(e^{\frac{3\Phi}{2}} F_{MN}\big) + \frac{1}{6}  e^{\frac{\Phi}{2}}  F_{PQRN} H^{PQR} \ , \cr 
0 &=& \nabla^M \big(e^{\frac{\Phi}{2}} F_{MNPQ}\big) - \frac{1}{144} \epsilon_{M_1 ... M_7 N P Q} F^{M_1 ... M_4} H^{M_5 ... M_8}  \ ,  
\end{eqnarray}
where $R_{MN}$ is the Ricci tensor and $\epsilon_{M_1 ... M_{10}}$ is totally antisymmetric tensor. 

Here, $\alpha_p$ is a $p$-form flux with $\alpha_p^2 = \alpha_{M_1...M_p}  \alpha^{M_1...M_p}$.
Taking the trace of the first equation in \eqref{EOM} and by plugging it into the second equation we have the simplified dilaton equation of the form
\begin{eqnarray}
\nabla^M \nabla_M \Phi - 2R + g^{MN} \partial_M \Phi \partial_N \Phi + \frac{1}{6} e^{-\Phi} \big(H_3\big)^2 = 0, 
\end{eqnarray}
where $R$ is the Ricci scalar. 

In addition to the bosonic sector, the massive Type-IIA supergravity also contains the gravitino $\Psi_M$ as well as the dilatino $\Lambda$ sector along with 32-component Majorana spinors. Their corresponding equations of motion are expressed as
\begin{eqnarray}
 0 &=& \Gamma^{MNP} D_N \Psi_P - \frac{1}{4} d\Phi \  .  \ \Gamma^M \Lambda + \frac{1}{4} F_0 e^{\frac{5\Phi}{4}}  \Gamma^{MN}  \Psi_N + \frac{5}{16} F_0 e^{\frac{5\Phi}{4}} \Gamma^M \Lambda  \cr
 && - \frac{1}{8} e^{\frac{3\Phi}{4}} \Bigg(2\Gamma^{[M]} F_2  \  .  \ \Gamma^{[N]}  \Gamma_{11} \Psi_N - \frac{3}{2} F_2  \  .  \ \Gamma^M \Gamma_{11} \Lambda\Bigg) \cr
 && - \frac{1}{8} e^{-\frac{\Phi}{2}} \Bigg(2\Gamma^{[M]} H_3  \  .  \ \Gamma^{[N]}  \Gamma_{11} \Psi_N -  H_3  \  .  \ \Gamma^M \Gamma_{11} \Lambda\Bigg) \cr
 && + \frac{1}{8} e^{\frac{\Phi}{4}} \Bigg(2\Gamma^{[M]} F_4  \  .  \ \Gamma^{[N]}   \Psi_N + \frac{1}{2}  F_4  \  .  \ \Gamma^M  \Lambda\Bigg),\\
 0 &=& \Gamma^M \nabla_M \Lambda - \frac{5}{16} e^{\frac{3\Phi}{4}} F_2  \  .  \ \Gamma_{11} \Lambda + \frac{3}{8} e^{\frac{3\Phi}{4}}  \Gamma^M  F_2   \  .  \ \Gamma_{11} \Psi_M \cr
 && + \frac{1}{4} e^{-\frac{\Phi}{2}} \Gamma^M H_3  \  .  \ \Gamma_{11}  \Psi_M +  \frac{3}{16} e^{\frac{\Phi}{4}} F_4  \  .  \ \Lambda - \frac{1}{8} e^{\frac{\Phi}{4}} \Gamma^M F_4  \  .  \ \Psi_M \cr
 && - \frac{1}{2} \Gamma^M d\Phi  \  .  \ \Psi_M - \frac{21}{16} F_0 e^{\frac{5\Phi}{4}} \Lambda - \frac{5}{8} F_0 e^{\frac{5\Phi}{4}} \Gamma^M \Psi_M . 
\end{eqnarray}

Here $\nabla_M$ is the spin-covariant derivative that acts on spinors and $.$ denotes Clifford product $\alpha_p  \ . \ \Lambda = \frac{1}{p !} \alpha_{M_1 .. M_P} \Gamma^{M_1 .. M_P} \Lambda$.
The matrices $\Gamma_M$ generate the Clifford algebra $CL(1,9)$ with $\{\Gamma^M , \Gamma^N\} = 2 g^{MN}$.
The chirality operator is defined as $\Gamma_{11} = \Gamma_0 ... \Gamma_9$. 
\section{Fluctuations of the Type IIA background}
Below, we consider the fluctuations of the supergravity fields as 
\begin{eqnarray} \label{per}
g_{MN} &=&  \bar{g}_{MN} +h_{MN},\\
\Phi &=& \bar{\Phi} + \varphi ,\\
H_3 &=& \bar{H}_3 + \delta H_3,\\
F_2 &=& \bar{F}_2 + \delta F_2,\\
F_4 &=& \bar{F}_4 + \delta F_4,
 \end{eqnarray}
which yields the $R_{MN}$ equation at leading order as 
\begin{eqnarray} \label{Rh} 
0 &=&\frac{1}{2} \tilde{\nabla}^B  \tilde{\nabla}_M h_{BN}  + \frac{1}{2} \tilde{\nabla}^{B} \tilde{\nabla}_N h_{B M}  - \frac{1}{2}  \tilde{\nabla}^2 h_{MN} - \frac{1}{2}  \tilde{\nabla}_N  \tilde{\nabla}_M \tilde{h}  + 4 \tilde{\nabla}^{B} A \tilde{\nabla}_{M} h_{BN} \cr
&& + 4 \tilde{\nabla}^{B} A \tilde{\nabla}_{N} h_{BM} - h_{MN} \tilde{\nabla}^2 A - 8 h_{MN} \big( \tilde{\nabla} A\big)^2 -  4 \tilde{\nabla}^{P} A \tilde{\nabla}_{P} h_{MN}  - \frac{1}{2} \partial_M \varphi \partial_N \bar{\Phi} \cr
&&  - \frac{1}{2} \partial_M \bar{\Phi} \partial_N \varphi  - \frac{1}{2} \sum_{p=2}^{4} \gamma_p e^{2(1-p) A +\alpha_p \bar{\Phi}} \Bigg[\big(\delta \mathcal A_p\big)_M \ .  \  \big( \tilde {\mathcal A_p}\big)_N +  \big( \tilde {\mathcal A_p}\big)_M  \ .  \  \big(\delta \mathcal A_p\big)_N \cr
&& - \beta_p h_{MN} \tilde {\mathcal A_p}^2 - \big(p-1\big) h_{PK}  \tilde {\mathcal A}_{M A_1 .. A_{p-2}}^P   \tilde {\mathcal A}_{N}^{A_1 .. A_{p-2} K}\Bigg] - \frac{\varphi}{2} \sum_{p=2}^{4} \alpha_p \gamma_p e^{2(1-p) A +\alpha_p \bar{\Phi}} \big(\tilde {\mathcal A_p}^2\big)_{MN} \cr
&& +t \tilde{g}_{MN}. 
\end{eqnarray}

Here, we define the above function as
\begin{eqnarray} 
\label{t}
t  &=&  \tilde{\nabla}^B  h_{BP} \tilde{\nabla}^P A + h_{BP} \tilde{\nabla}^B \tilde{\nabla}^P A - \frac{1}{2} \tilde{\nabla}_{\Lambda} \tilde{h}  \tilde{\nabla}^{\Lambda} A + 8 h_{PB} \tilde{\nabla}^{P} A \tilde{\nabla}^{B} A  \cr
&& + \frac{\varphi}{2} \sum_{p=2}^{4} \alpha_p \beta_p \gamma_p e^{2(1-p) A +\alpha_p \bar{\Phi}} \tilde {\mathcal A_p}^2 + \frac{1}{2}  \sum_{p=2}^{4} \beta_p \gamma_p e^{2(1-p) A +\alpha_p \bar{\Phi}} \cr
&&\Bigg[2 \big(\delta \mathcal A_p\big)  \ .  \  \big( \tilde {\mathcal A_p}\big) - p h_{PK} \tilde {\mathcal A}_{ A_1 .. A_{p-1}}^P   \tilde {\mathcal A}_{N}^{A_1 .. A_{p-1} K}\Bigg] \ , 
\end{eqnarray}
where $\tilde{h} = \tilde{g}^{MN} h_{MN}$ . 
\section{Adding flavour branes}\label{appen c}
In this Appendix, we discuss effects of adding flavour D6 branes into the above picture and the associated massive graviton equations. As we show below, these equations are quite involved and almost impossible to solve analytically. The purpose of the section is to outline the basic structure of these equations which may be useful for future investigations. 

We begin by considering the single kink profile whose corresponding potential function is expressed as \cite{Roychowdhury:2021eas}
\begin{eqnarray} 
V (\sigma \sim  0, \eta) = \eta N_6 \ln \sigma + \frac{\eta N_6 \sigma^2}{4} \Lambda_k (\eta, P) -  \frac{\eta N_6 \sigma^2}{4}   \frac{P+1}{P^2 - \eta^2}, 
\end{eqnarray}
where we define
\begin{eqnarray} 
\Lambda_k (\eta, P) &=& \big(P+1\big) \sum_{m = 1}^k \Bigg[\frac{1}{(2m+(2m-1)P)^2 - \eta^2} -  \frac{1}{(2m+(2m+1)P)^2 - \eta^2}\Bigg] \cr
&& + \frac{P}{(2k+1)^2(1+P)^2 - \eta^2}. 
\end{eqnarray}

The associated metric functions can be expressed as
\begin{eqnarray} \label{f fla}
f_1 (\sigma \sim 0, \eta) &\sim& \frac{4}{\sigma} \frac{1}{g(\eta)}, \\
f_2 (\sigma \sim 0, \eta) &\sim& \sigma g(\eta),\\
f_3 (\sigma \sim 0, \eta) &\sim& \eta^2 \sigma g(\eta),\\
f_4 (\sigma \sim 0, \eta) &\sim& 0, 
\end{eqnarray}
along with the function
\begin{eqnarray} \label{g} 
g(\eta) = \frac{\sqrt{2}}{\sqrt{ \eta} (P^2 - \eta^2)^{\frac{3}{2}}} \Bigg[\big(P^2 - n^2\big)^3  \big(\eta \partial_\eta^2 \Lambda + 2 \partial_\eta \Lambda\big) - 2 \eta \big(P+1\big) \big(\eta^2 + 3P^2\big)\Bigg]^{\frac{1}{2}}. 
\end{eqnarray}

Using the above information \eqref{f fla}-\eqref{g}, it is straightforward to compute
\begin{eqnarray} \label{L - flav}
 \mathcal L^{(1)} F &=& \frac{1}{\eta^2 \sigma^3 g(\eta)^3}  16 \Bigg[g(\eta) \Big(\sigma \Big(\eta \Big(-2 \eta \partial_\sigma \Phi \partial_\sigma F  - 2 \Big(\eta \partial_\eta \Phi -1\Big) \partial_\eta F + \eta \partial_\sigma^2 F \cr 
 && + \eta  \partial_\eta^2 F\Big) + \partial_\chi F \Big(\cot \chi - 2 \partial_\chi \Phi \Big)+ \csc^2 \chi \Big(\partial_\xi^2 F - 2  \partial_\xi \Phi  \partial_\xi F\Big) + \partial_\chi^2 F \Big) \cr
 && - 2 \eta^2 \partial_\sigma F \Big) - 2 \eta^2 \sigma \partial_\eta g(\eta) \partial_\eta F\Bigg]. 
 \end{eqnarray}

Next, we consider the Uluru profile, for which the potential function is expressed as \cite{Roychowdhury:2021eas}
\begin{eqnarray} 
V (\sigma \sim  0, \eta) = - \eta N_6 \ln \sigma + \frac{\eta N_6 \sigma^2}{4} \Lambda_u (\eta, K, P) +  \frac{\eta N_6 \sigma^2}{4 (P^2 - \eta^2)}, 
\end{eqnarray}
where we define
\begin{eqnarray} 
\Lambda_u (\eta, K, P) = \sum_{n = 1}^u (-1)^{n+1}  \Bigg[\frac{1}{(nK+(2n-1)P)^2 - \eta^2} -  \frac{1}{(nK+(2n+1)P)^2 - \eta^2}\Bigg]. 
 \end{eqnarray}

This results in a similar equation \eqref{L - flav} together with a function
\begin{eqnarray} 
g(\eta) = \frac{\sqrt{2}}{\sqrt{ \eta} (\eta^2 - P^2)^{\frac{3}{2}}} \Bigg[2\eta^3 + \big(P^2 - n^2\big)^3  \big(\eta \partial_\eta^2 \Lambda_u + 2 \partial_\eta \Lambda_u\big) + 6 \eta P^2 \Bigg]^{\frac{1}{2}}.
\end{eqnarray}

\end{document}